# Effect of Co-Ga paired substitution on superconductivity in $YBa_2Cu_3O_{7-\delta}$

Anjana Dogra[1], S. Rayaprol[2], N. A. Shah[3] and D. G. Kuberkar[4]

*1. Material Science Division, Nuclear Science Centre, New Delhi – 110 067*

*2. DCMP & MS, Tata Institute of Fundamental Research, Mumbai – 400 005*

*3. Department of Electronics, Saurashtra University, Rajkot – 360 005*

*4. Department of Physics, Saurashtra University, Rajkot – 360 005*

Corresponding Author

Dr. S. Rayaprol

DCMP & MS, Tata Institute of Fundamental Research

Homi Bhabha Road, Colaba

Mumbai – 400 005, INDIA

Email: sudhindra@rayaprol.com

Fax : +91-22-2280 4610

**ABSTRACT**

The effect of Co-Ga paired substitution on the superconducting properties of $YBa_2Cu_3O_{7-\delta}$ (Y-123) has been investigated by X-ray diffraction, ac susceptibility, dc resistivity and oxygen content measurements. We report in this paper the results of our studies on the paired substitution of a magnetic and non-magnetic ion at Cu site in Y-123, while keeping the total dopant concentration fixed. The simultaneous substitution of Co and Ga at Cu site shows variation in the transition temperature ($T_c$), oxygen content and hole concentration as a function of change in the balance of magnetic (Co) and non-magnetic (Ga) concentration. Orthorhombicity (D), given as (b-a)/a, also varies as a function of increasing dopant concentration. The variation in $T_c$ due to Co-Ga substitution is discussed in the light of dopant valency and hole filling mechanism.

## 1. INTRODUCTION

Substitutions at various crystallographic sites in the triple perovskite Y-123 structure has been a very popular approach of understanding the mechanism of superconductivity in these high temperature superconductors (HTSC) [1, 2]. The substitution of rare earth (RE) ion at Y site yields isostructural RE-123 superconducting phases (except for RE = Ce, Tb and Pr) with $T_c \sim 90$ K [3]. Substitution of alkaline earth metals (e.g., Ca or Sr) at Ba site leads to suppression of $T_c$ [4]. But the substitution of Cu site by transition metal (e.g., Co, Fe, Ni, Zn etc) or non-transition metal element (e.g., Ga and Al) always results into $T_c$ suppression [5, 6].

Superconductivity in Y-123 triple perovskite structures resides in the Cu-O sheets. Y-123 structure has crystallographically two distinct Cu sites, Cu(1) site in CuO chains and Cu(2) in $CuO_2$ sheets. It is known that $CuO_2$ layers play an important role in transferring of the charge carriers, where as CuO chains are non-superconducting and acts as 'charge reservoir'. From the structural point of view, substitution at the Cu(1) site can cause increase in the oxygen content in the basal plane, inducing orthorhombic to tetragonal (O-T) transition with increasing dopant concentration. If the substitution takes places at Cu(2) site, there is no observed structural O-T transition, the structure remains orthorhombic even for higher dopant concentrations and oxygen order in the CuO chains remains intact [7].

The divalent cations substitute preferentially at the Cu(2) site where as trivalent cations substitute preferentially at the Cu(1) site. 3d- transition elements (like Co, Ni and Fe) have a specific preference for the Cu-site in the Y-123 system [8]. The $Co^{3+}$ ion prefers distinctively the Cu(1) site whereas the $Ni^{3+}$ ion chooses the Cu(2) site but Fe ion is known to take any of the two Cu sites [9]. In order to elucidate the role of the Cu-sites in determining the superconducting properties, the ideal dopants are the non-magnetic elements with electronic configuration similar to that of $Cu^{2+/3+}$ and which should preferentially substitute at either Cu(1) or at the Cu(2) site. $Ga^{3+}$, though trivalent ($3d^{10}$) cation, has a closed shell and fixed valency and non-magnetic nature, unlike the 3d transition metal ions which exhibit multiple valencies (2+ / 3+ / 4+) [7, 10].

Many theoretical papers supported by experimental results have considered either magnetic transition metal ion (e.g., Co, Ni, Fe etc) or non-magnetic transition metal ion substitution (e.g., Gd, Zn, Al etc) for understanding the role of substitutional impurities in modifying the superconducting properties [11]. The theoretical developments have indicated that the antiferromagnetic spin fluctuations in HTSC favors *d-wave* pairing. But due to disorder in the structure of the orthorhombic HTSC, distortion of the pairing interaction takes places, giving rise to *s-wave* pairing component in the original *d-wave* pairing. Many experiments have shown that, the $CuO_2$ sheet in the Y-123 structure is a superconducting layer (S) where as the CuO chain layer is a non-superconductor (N) [12]. The hopping interaction between these S-N layers leads to the suppression of $T_c$ [13, 14].

To the best of our knowledge very few reports exists on the simultaneous substitution of a magnetic and non-magnetic metal ion at Cu site keeping the total dopant concentration fixed. In this paper we present the results of structural and superconducting property measurements on $YBa_2Cu_{2.88}(Co_{1-x}Ga_x)_{0.12}O_{7-\delta}$ for $0.0 < x < 1.0$ (thus referred as YCGO series). The results of X-ray diffraction, ac susceptibility and oxygen content measurement by iodometric double titration are discussed in detail. Since both $Co^{2+}$ and $Ga^{3+}$ are known to prefer Cu(1) site, it is interesting to vary the level of the magnetic and non-magnetic concentration and study its effect on the structural and superconducting properties.

## 2. EXPERIMENTAL DETAILS

All the samples of the YCGO series were synthesized by the standard solid state reaction method. The starting compounds of $Y_2O_3$, $BaCO_3$, CuO, $Co_3O_4$ and $Ga_2O_3$ (all > 99.9 % pure) were thoroughly mixed using agate and mortar under acetone and initially fired at $900^0$C for 24 hours in powder form. The resultant black powder was pressed in to cylindrical pellets and subjected to further heat treatment at $950^0$C for 48 hours in air with intermittent grindings in between. The palletized samples were then annealed in flowing oxygen at $500^0$C for about 24 hours and then slow cooled to room temperature. Finely powdered samples were taken up for X-ray diffraction using Cu-$K_\alpha$ radiation. All the samples exhibited single phase formation. Lattice parameters were calculated using the least square method. Superconducting transition temperatures were observed by ac susceptibility, which was cross checked by four probe dc resistance measurement. Double coil (mutual inductance) ac susceptibility setup operating at $\omega$ = 100Hz, was used to measure the diamagnetic signal on EG&G (PAR 5210) lock-in amplifier. Four probe resistivity setup, using an APD cryostat, was used to measure the change in resistance as function of temperature, as a cross check for the $T_c$ observed by susceptibility method. The transition temperatures measured by both the methods were in good agreement. The oxygen content and hole concentration per unit cell was determined by the Iodometric double titration method as described by Nazzal et.al [15].

## 3. RESULTS AND DISCUSSION

### *3.1 X-ray diffraction*

The variation in unit cell parameters as a function of dopant concentration is plotted in Figure 1. There is a marginal decrease in average value of 'a' and 'b' parameters, while c increases from 11.63Å to 11.67Å with x increasing from 0.0 to 1.0. With increasing Ga and decreasing Co at Cu site, there is an increase in the c-parameter, which can be attributed to the ionic size effect.

### *3.2 Resistivity and susceptibility*

Figure 2(a) shows the ac susceptibility plots for all the samples studied. Figure 2(b) shows the normalized resistance versus temperature plots for x = 0.0, 0.4 & 1.0. It can be seen from the figure that $T_c$ values of samples with x = 0.0 and 1.0 (only Co or Ga substituted samples respectively) agree with the values reported in the literature [8, 16]. With increasing x, $T_c$ is found to decrease upto ~ 50 K for x = 0.4 (i.e., higher Co values, compared to Ga). Above x = 0.4 when $Ga^{3+}$ content increases than $Co^{3+}$, the $T_c$ approaches 88 K as if there is only marginal effect of $Ga^{3+}$ substitution on superconductivity of Y-123. The susceptibility plot of sample x = 0.6, shows a hump in the diamagnetic region presumably due to some magnetic impurity phase.

### *3.3 Hole concentration and oxygen content variation*

A. I. Nazzal et al have given a procedure for determining hole concentration by determining the oxidation state of copper [15]. The method described by Nazzal et al has been carefully followed in determining the oxidation state of the copper by the iodometric titration of the powdered samples.

The following reaction occurs when the compound containing copper oxide is dissolved in an acidic solution with excess potassium Iodide (KI):

$$[Cu-O]^{+p} + (2+p)I^- \rightarrow CuI + \frac{(p+1)}{2} I_2 \quad \text{---(1)}$$

The sample, which contains copper oxide, is dissolved in an acidic solution. The following relation converts the copper present in the sample into $Cu^{2+}$:

$$Cu^+ + H^+ \xrightarrow{O} Cu^{2+} + H_2O \quad \text{---(2)}$$

$$[Cu-O]^{+p} \xrightarrow{H_2O} [Cu-O] + O_2 \quad \text{---(3)}$$

The degree of the oxidation or average charge of $[Cu\text{-}O]^{+p}$ is given by the relation,

$$p = \frac{\left(V_1 / W_1\right)}{\left(V_2 / W_2\right)} - 1 \quad \text{---(4)}$$

where V1 and V2 are the volume of Sodium Thiosulphate ($Na_2S_2O_2$) needed to titrate Iodine generated by two samples weighing $W_1$ and $W_2$ respectively. The formal copper valence is (2+p) and can be calculated if the stoichiometry is know, using charge neutrality.

For Y-123 compounds ($YBa_2Cu_3O_z$), according to the charge neutrality, oxygen content (z) can be calculated as

$$2z = 3(2+p) + 7 \quad \text{---(5)}$$

Figure 3 (a) shows the decrease in oxygen content with increasing dopant concentration. The hole concentration, calculated from the oxygen content per unit cell using the Equations (4) and (5), is also found to decrease with increasing dopant concentration, as can be seen in Figure 3 (b). The decrease in the hole concentration (p) can be attributed to the hole filling by $Co^{3+}$ or $Ga^{3+}$ or both at the Cu site. We have calculated the anisotropy factor (ANSI), which is given by the relation, ANSI = [100(b-a)/0.5(b+a)]. The ANSI gives the percentage deviation from tetragonality. For a typical, fully oxygenated Y-123 sample with z = 6.9, the ANSI factor is around 1.55, which exhibits orthogonal structure [2]. ANSI exhibits a shift towards tetragonality for x = 0.4, 0.6 and 0.8.

The structural transition can also be observed by calculating the orthorhombicity of the unit cell. Figure 4 (a) shows the variation of orthorhombicity with increasing dopant concentration. Transition temperature also varies in a similar fashion as shown in Figure 4 (b). There is drop in both orthorhombicity and $T_c$ upto x = 0.4, and then increases with increasing x. The rate of change of $T_c$ with x, given by $dT_c/dx$ is shown in Figure 4 (c).

From Figures 3 and 4, we can say that there is a strong correlation between the structure and superconducting transition temperature for these oxides. For samples between x = 0.0 and 1.0 the structural and superconducting properties are governed by competition between magnetic ion and non-magnetic ion substitution. The substitution of $Co^{3+}$ and $Ga^{3+}$ at the Cu(1) site results in the combined effect on the superconducting transition properties.

## 4. CONCLUSIONS

We have presented here the results of combined effect of a magnetic ($Co^{3+}$) and non-magnetic ($Ga^{3+}$) element substitution at Cu site, on structure, superconducting transition temperature, oxygen content, hole concentration of a Y-123 system. The observed non-systematic variation in $T_c$ and orthorhombicity may be attributed to the combined effect of Co and Ga in modifying the superconducting properties. Assuming that valency of dopants is fixed, we see that the additional electrons contributed by $Co^{3+}$ and $Ga^{3+}$ at $Cu^{1+}$ and $Cu^{2+}$ sites respectively results in the reduction of $T_c$. It has been observed that though oxygen content and hole concentration decreases with increase in dopant concentration, there has been restoration of superconducting transition temperature. The distribution of Co and Ga at Cu(1) and Cu(2) sites, due to their site preferences, as discussed earlier is assumed to be responsible for such behavior. We have also observed that when the concentration of magnetic dopant is higher at the Cu-site, the $T_c$ is suppressed faster than the non-magnetic dopant for the same concentration. This information highlights the role played by the moment of the magnetic ion in influencing the superconducting properties.

More experiments and theoretical considerations on these complex systems, considering both magnetic and non-magnetic ion substitution at Cu site, would be more helpful in understanding the role of CuO layers in structure and superconductivity of Y-123 systems.

ACKNOWLEDGEMENTS

Authors wish to thank Prof. R. G. Kulkarni for fruitful discussions and to Dr. R. S. Thampi for electrical measurements and XRD results.

FIGURE CAPTION

Figure 1 Unit cell parameters a, b, c (all in Å) as a function of dopant concentration.

Figure 2 (a) ac susceptibility ($\chi_{ac}$) as a function of temperature for all samples in the $YBa_2Cu_{2.88}(Co_{1-x}Ga_x)_{0.12}O_{7-\delta}$ series.

Figure 2(b) Typical normalized resistance for x = 0.0, 0.4 and 1.0 samples of $YBa_2Cu_{2.88}(Co_{1-x}Ga_x)_{0.12}O_{7-\delta}$ series as a function of temperature. The resistance values measured was of the order of milliohms.

Figure 3 (a) The oxygen content and (b) hole concentration, per formula unit of $YBa_2Cu_{2.88}(Co_{1-x}Ga_x)_{0.12}O_{7-\delta}$ series. (c) The anisotropic factor, ANSI given as (100*(b-a)/0.5*(b+a)) varies with changes in the Co and Ga concentration at Cu site.

Figure 4 Change in

(a) Orthorhombicity (D), given as (b-a)/a

(b) transition temperature ($T_c$), inferred from $\chi_{ac}$ measurements

(c) rate of change of $T_c$ with dopant concentration

for $YBa_2Cu_{2.88}(Co_{1-x}Ga_x)_{0.12}O_{7-\delta}$ series

Figure 1

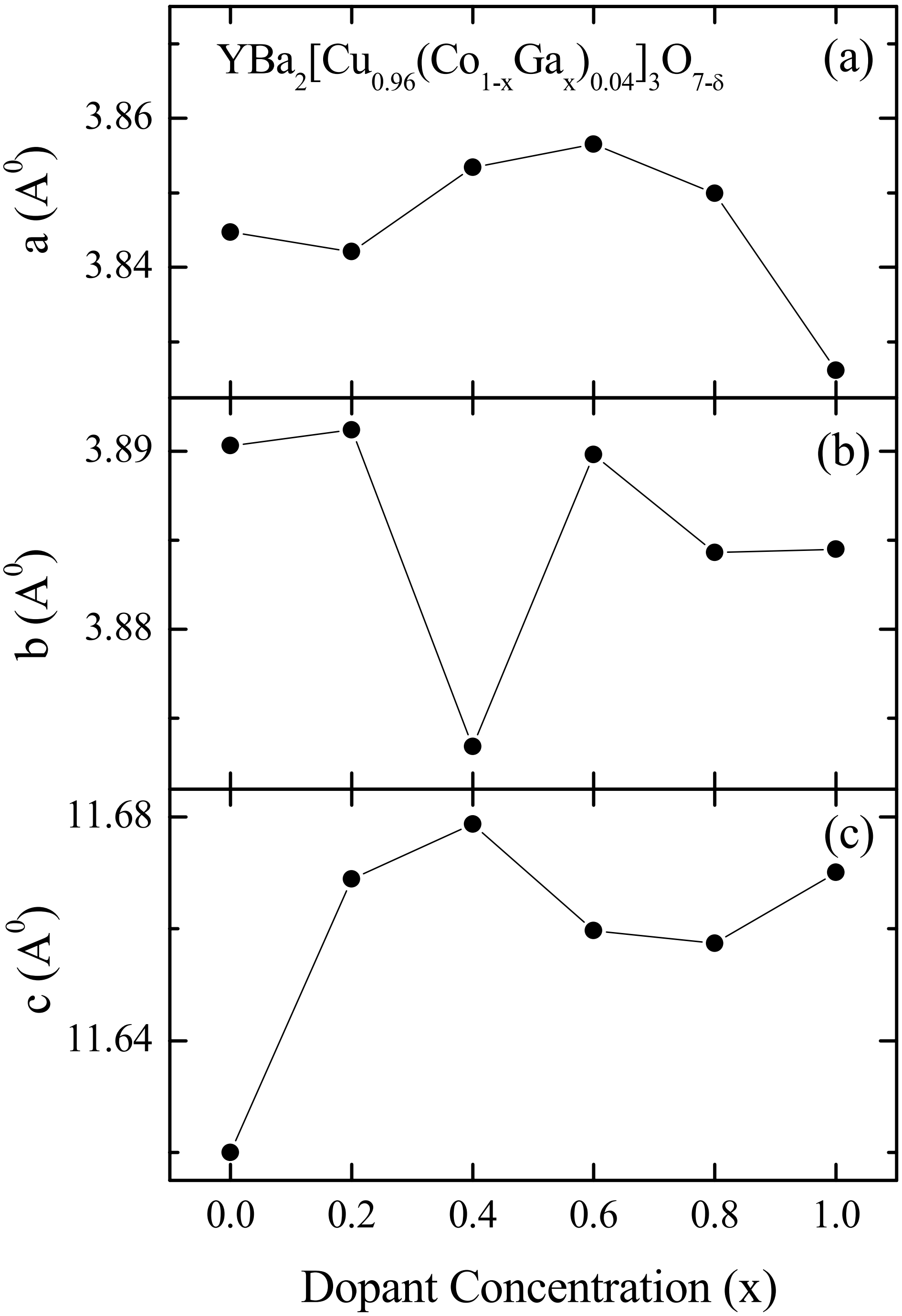

Figure 2 (a)

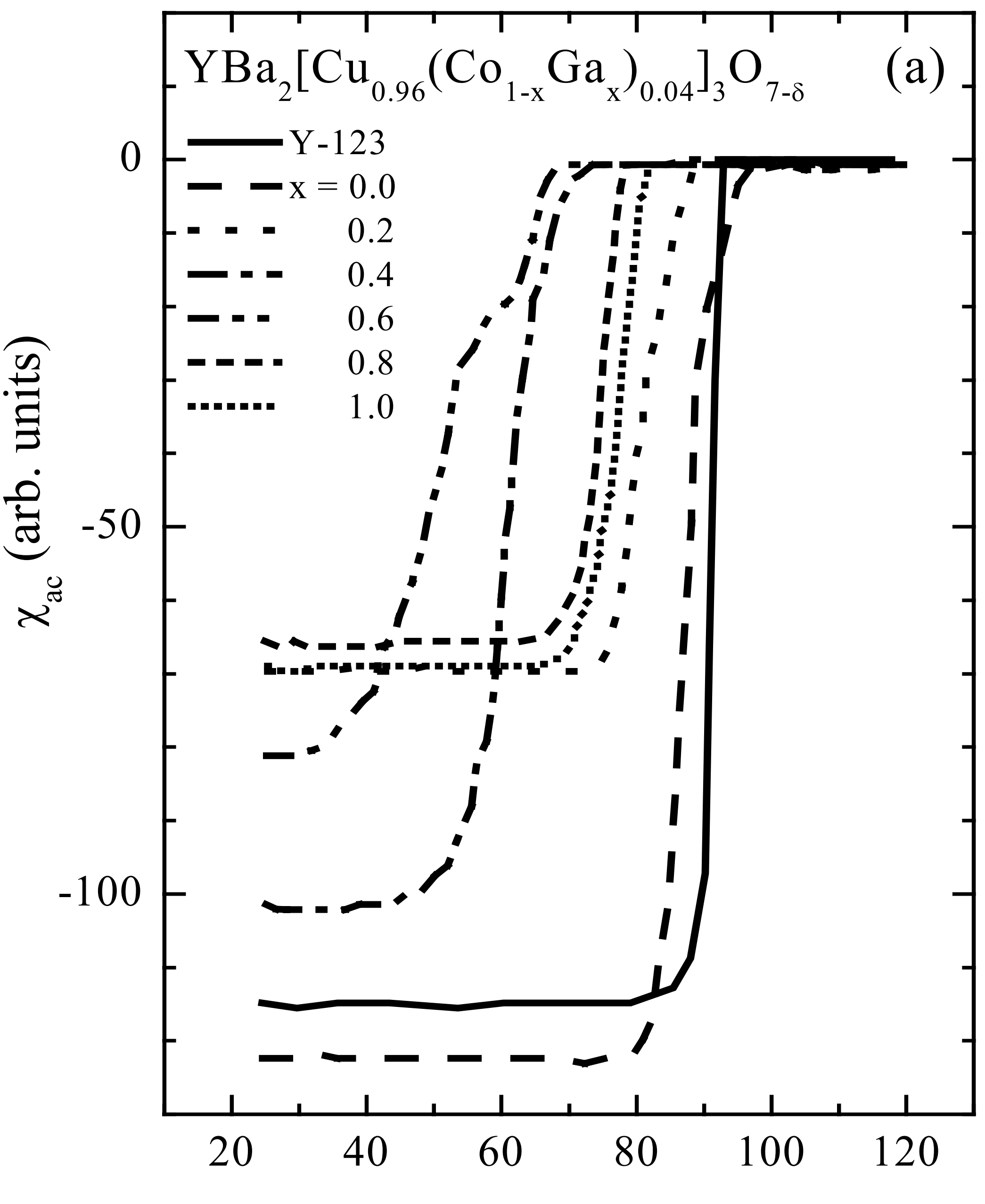
YBa$_2$[Cu$_{0.96}$(Co$_{1-x}$Ga$_x$)$_{0.04}$]$_3$O$_{7-\delta}$
(a)
Y-123
x = 0.0
0.2
0.4
0.6
0.8
1.0
0
-50
-100
20
40
60
80
100
120
$\chi_{ac}$ (arb. units)
T (K)

Figure 2 (b)

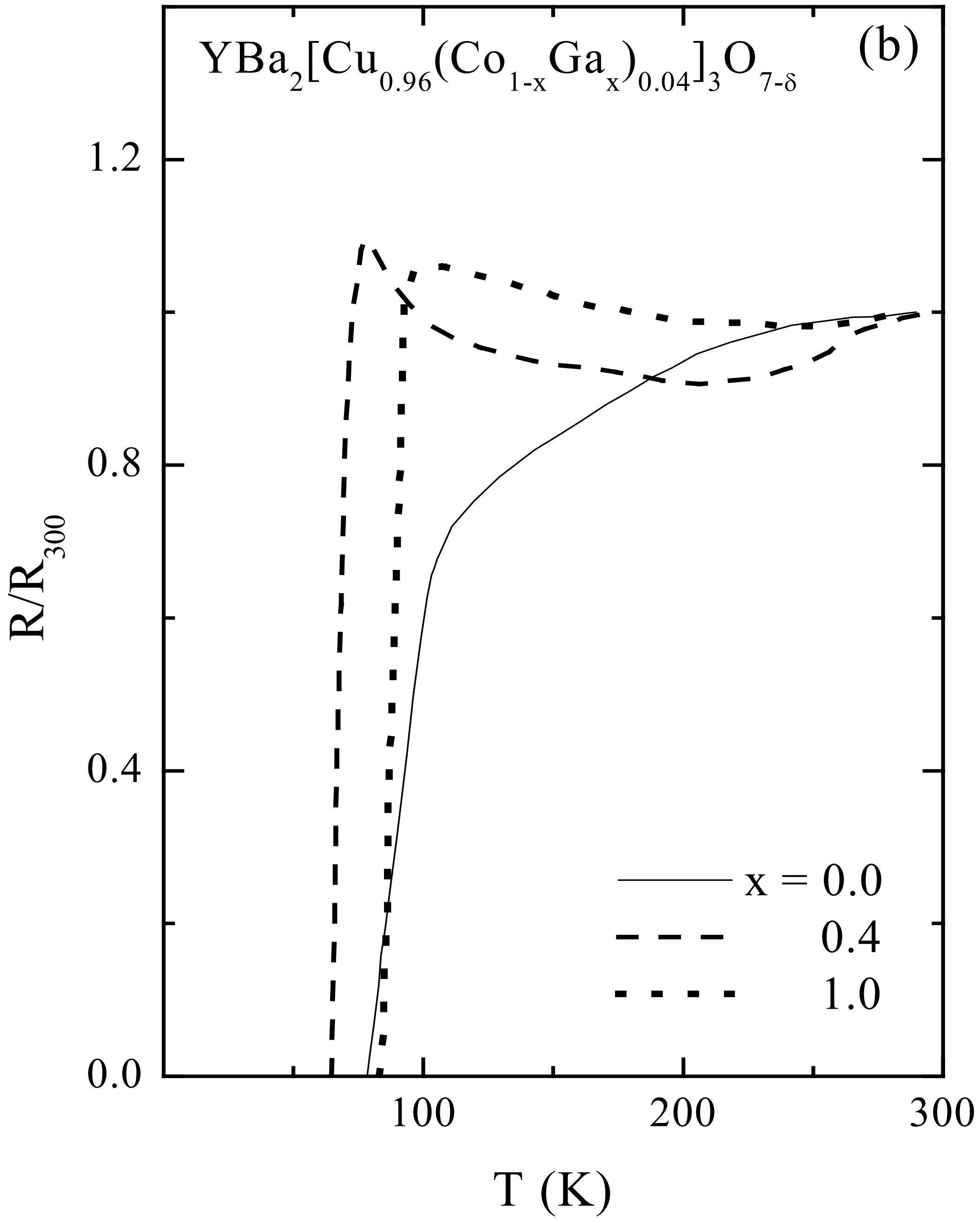

Figure 3

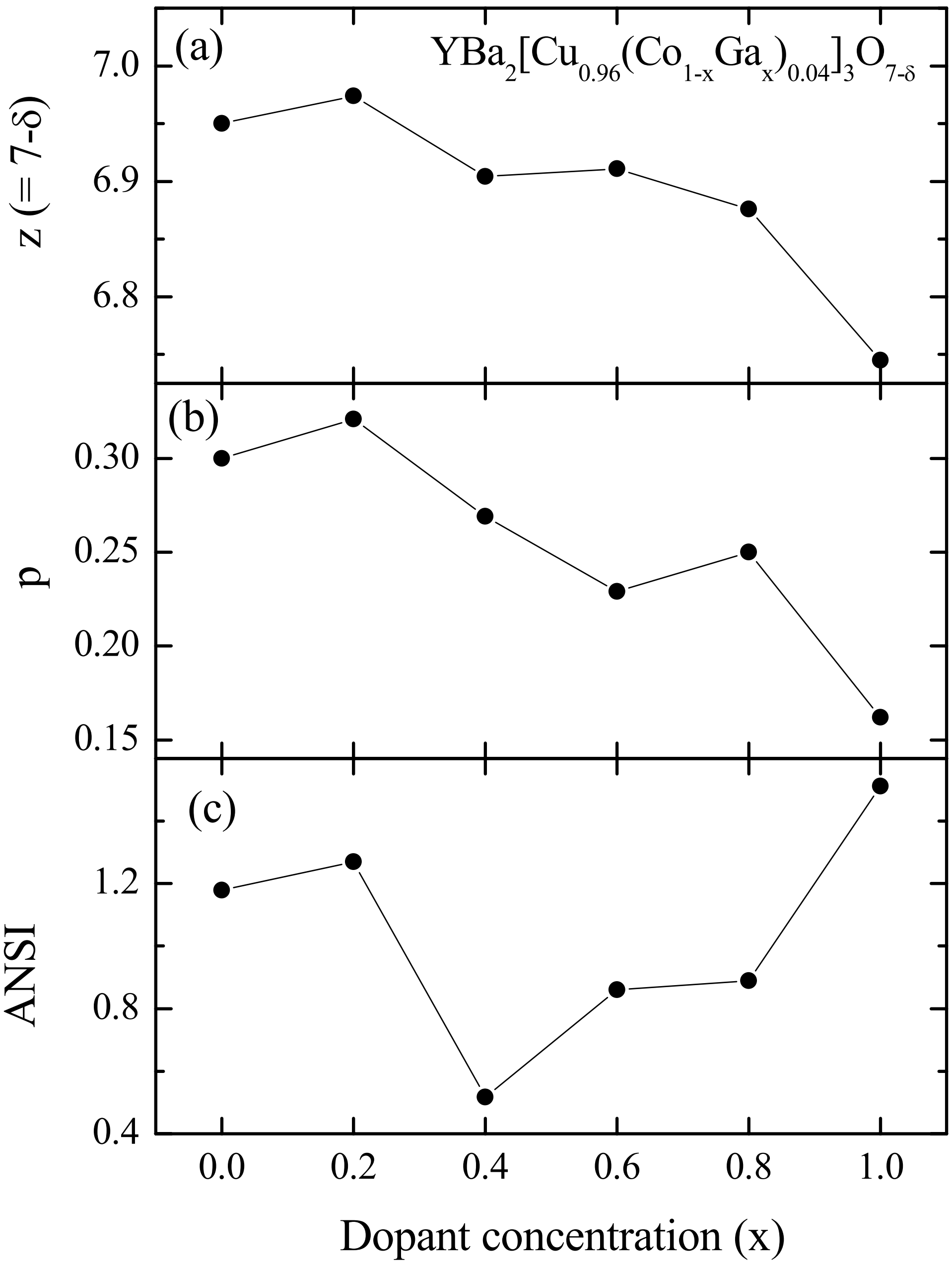
(a)
YBa$_2$[Cu$_{0.96}$(Co$_{1-x}$Ga$_x$)$_{0.04}$]$_3$O$_{7-\delta}$
z (= 7-δ)
7.0
6.9
6.8
(b)
p
0.30
0.25
0.20
0.15
(c)
ANSI
1.2
0.8
0.4
0.0
0.2
0.4
0.6
0.8
1.0
Dopant concentration (x)

Figure 4

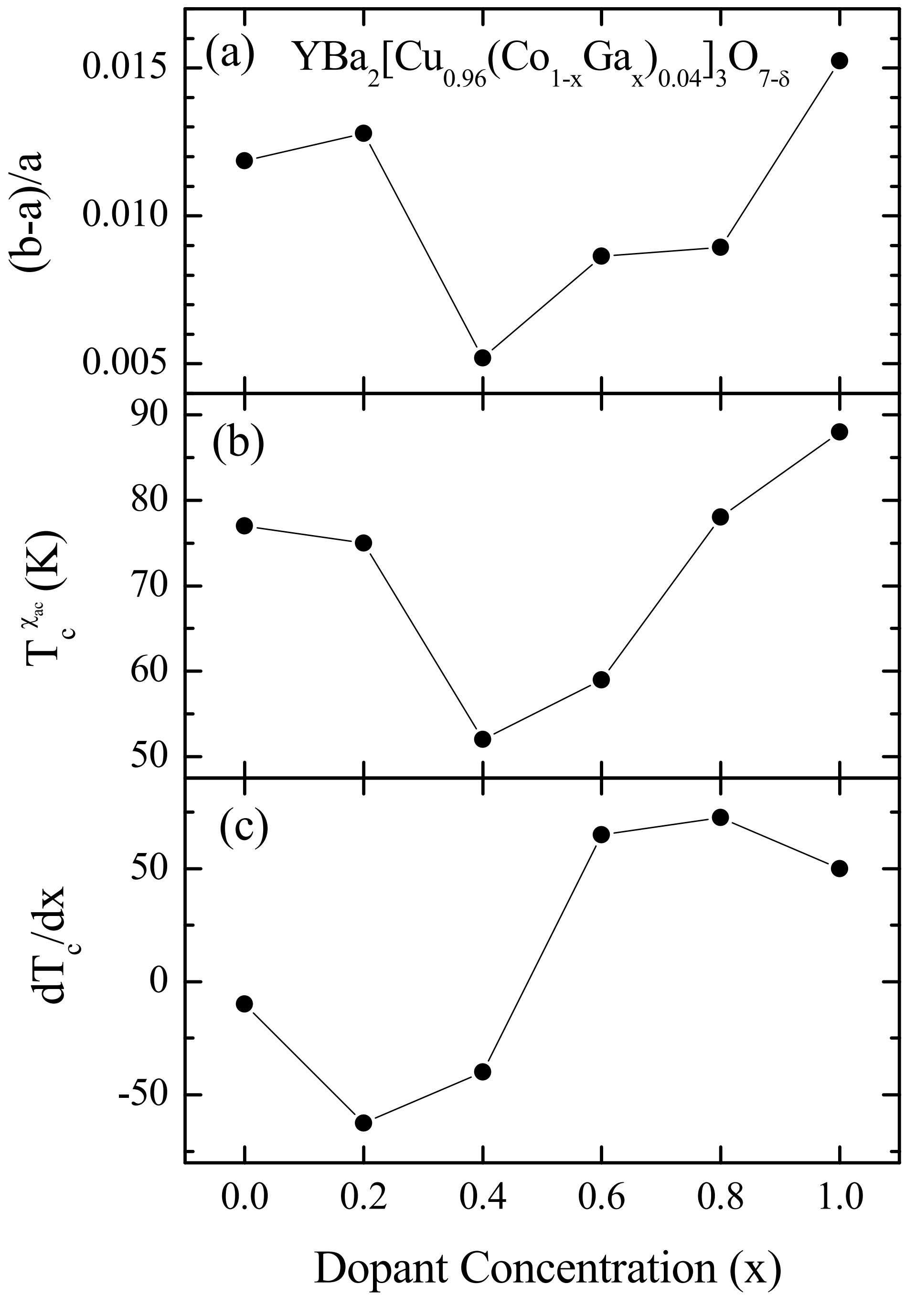